\documentclass[12pt]{article}
\usepackage{epsfig,amssymb,amsmath,psfrag,subfigure}

\newcommand{\insertfig}[2]{\mbox{\epsfxsize=#1cm \epsfbox{#2.eps}}}

\def \be  {\begin{equation}}
\def \ee  {\end{equation}}
\def \ba  {\begin{eqnarray}}
\def \ea  {\end{eqnarray}}
\def \baa {\begin{eqnarray*}}
\def \eaa {\end{eqnarray*}}
\def \lab #1 {\label{#1}}

\newcommand\re[1]{(\ref{#1})}

\def \matrix #1 {\left(\begin{array}{cc} #1 \end{array}\right)}

\def \tr {\mathop{\rm tr}\nolimits}

\newcommand \vev [1] {\langle{#1}\rangle}
\newcommand \VEV [1] {\left\langle{#1}\right\rangle}
\newcommand \ket [1] {|{#1}\rangle}
\newcommand \bra [1] {\langle {#1}|}

\newcommand{\ft}[2]{{\textstyle\frac{#1}{#2}}}
\begin{document}

\begin{titlepage}

\thispagestyle{empty}

\vspace*{3cm}

\centerline{\large \bf Conformal anomaly of super Wilson loop}
\vspace*{1cm}

\centerline{\sc A.V. Belitsky}

\vspace{10mm}

\centerline{\it Department of Physics, Arizona State University}
\centerline{\it Tempe, AZ 85287-1504, USA}

\vspace{1cm}

\centerline{\bf Abstract}

\vspace{5mm}

Classically supersymmetric Wilson loop on a null polygonal contour possesses all symmetries required to match it onto non-MHV amplitudes 
in maximally supersymmetric Yang-Mills theory. However, to define it quantum mechanically, one is forced to regularize it since perturbative 
loop diagrams are not well-defined due to presence of ultraviolet divergences stemming from integration in the vicinity of the cusps. A regularization 
that is adopted by practitioners by allowing one to use spinor helicity formalism, on the one hand, and systematically go to higher orders of perturbation 
theory is based on a version of dimensional regularization, known as Four-Dimensional Helicity scheme. Recently it was demonstrated that its use for 
the super Wilson loop at one loop breaks both conformal symmetry and Poincar\'e supersymmetry. Presently, we exhibit the origin for these effects and 
demonstrate how one can undo this breaking. The phenomenon is alike the one emerging in renormalization group mixing of conformal operators 
in conformal theories when one uses dimensional regularization. The rotation matrix to the diagonal basis is found by means of computing the anomaly 
in the Ward identity for the conformal boost. Presently, we apply this ideology to the super Wilson loop. We compute the one-loop conformal anomaly for 
the super Wilson loop and find that the anomaly depends on its Grassmann coordinates. By subtracting this anomalous contribution from the super Wilson 
loop we restore its interpretation as a dual description for reduced non-MHV amplitudes which are expressed in terms of superconformal invariants.
 
\end{titlepage}

\setcounter{footnote} 0

\newpage

\pagestyle{plain}
\setcounter{page} 1

\section{Introduction}

In the past decade, the planar maximally supersymmetric Yang-Mills (SYM) theory became a laboratory for developing and
testing various techniques for strong coupling analysis of four-dimensional gauge theories via the AdS/CFT
correspondence. These discoveries stretch far beyond consequences of the exact superconformal symmetry
of the theory valid to any loop order \cite{BriMan82}. The focus of the last several years was the space-time S-matrix of the 
regularized theory\footnote{Since the theory does not develop a mass gap, the four-dimensional scattering 
amplitudes vanish due to infrared divergences intrinsic to theories with massless gauge bosons.}.

Due to the fact that all propagating fields of the $\mathcal{N} = 4$ SYM can be combined into a single light-cone superfield
$\Phi$ \cite{BriMan82,Nai88}, the $n$-particle matrix element of the S-matrix can be organized in a single superamplitude $\mathcal{A}_n$ 
which is defined in turn by the amputated Green function $\vev{\Phi_1 \Phi_2 \dots \Phi_n}_{\rm amp}$. Making use of the translation
invariance and Poincar\'e supersymmetry, one can extract the (super)momentum conservation laws from the superamplitude
and write it in the following generic form \cite{Nai88}
\be
\label{GenericA}
\mathcal{A}_n 
= 
i (2 \pi)^4 
\frac{
\delta^{(4)} \big( \sum\nolimits_i \lambda_i \tilde\lambda_i \big)  
\delta^{(8)} \big( \sum\nolimits_i \lambda_i \eta_i \big) 
}{\vev{12} \vev{23} \vev{34} \dots \vev{n-1 n}}
\widehat{\mathcal{A}}_n (\lambda_i, \widetilde\lambda_i, \eta_i; a)
\, .
\ee
Here we introduced a notation for the 't Hooft coupling constant $a = g^2 N_c/(4 \pi^2)$ and used the spinor helicity formalism to represent 
the particles' momenta $p_i^{\dot\alpha\alpha} = \widetilde\lambda^{\dot\alpha} \lambda^\alpha$ and chiral charges $q^A_\alpha = \eta^A 
\lambda_\alpha$ in terms of commuting Weyl spinors $\lambda_\alpha$ and anticommuting Grassmann variables $\eta^A$. Here the angle 
brackets are conventionally defined as $\vev{ij} = \lambda^\alpha_i \lambda_{j \alpha}$. The dependence of the scattering amplitude on particle 
helicities involved in scattering emerges from the expansion of the reduced superamplitude $\widehat{\mathcal{A}}_n$ in terms of $\eta$'s. 
Namely, it admits the following series
\be
\widehat{\mathcal{A}}_n = \widehat{\mathcal{A}}_{n, 0} + \widehat{\mathcal{A}}_{n, 1} + \dots
\, ,
\ee
with each term being a homogeneous polynomial of degree $\eta^{4 k}$. This expansion terminates at the term of order $k=n-4$ due to nilpotence of 
Grassmann variables and superconformal symmetry. The leading term of the reduced amplitude starts with unity at zero order 
in coupling since the maximal helicity-violating (MHV) amplitude was extracted from $\mathcal{A}_n$ in the form of the Parke-Taylor tree formula \cite{ParTay86}. 
The subleading terms are identified with N${}^k$MHV amplitudes, $\widehat{\mathcal{A}}_{n, k} = \widehat{\mathcal{A}}{\,}_n^{{\rm N}{}^k \rm{MHV}}$. 
While the tree superamplitude enjoys the full superconformal symmetry of the theory, one violates a number of them as quantum effects are 
taken into account, such that the dilatation, conformal boost and superconformal transformations are broken \cite{Wit04} by the infrared diverges due to copious 
emissions of particles.

As was discovered more recently \cite{DruHenKorSok07}, if the tree superamplitude is rewritten in terms of the so-called region supermomenta 
$(x_i, \theta^A_i)$, which solve automatically the supermomentum conservation conditions in  Eq.\ \re{GenericA},
\be
\widetilde\lambda^{\dot\alpha}_i \lambda^\alpha_i = x_{i i+1}^{\dot\alpha\alpha}
\, , \qquad
\lambda^\alpha_i \eta^A_i = \theta_{ii+1}^{\alpha A}
\, ,
\ee
the former exhibits yet another symmetry. The latter is the so-called dual superconformal symmetry which acts on $(x_i, \theta^A_i)$ as if they
were coordinates (not momenta). Making use of these coordinates, the tree-level NMHV amplitudes can be written in terms of superconformal 
invariants as follows \cite{DruHenKorSok07}
\be
\label{treeNMHV}
\widehat{\mathcal{A}}_{n, 1} =  \sum_{1< q < r < n} R_{nqr}
\, ,
\ee
where
\be
R_{pqr} = - 
\frac{
\vev{q-1 q} \vev{t-1 t} \delta^{(4)} (\Xi_{pqr})
}{
x_{qr}^2 \bra{r} x_{pr} x_{rq} \ket{q-1} \bra{p} x_{pr} x_{rq} \ket{q} \bra{p} x_{pq} x_{qr} \ket{r-1} \bra{p} x_{pq} x_{qr} \ket{r}
}
\, ,
\ee
and the argument of the Grassmann delta function is given by the expression
\be
\Xi_{pqr} = \sum_{i = r}^{p - 1} \eta_i \bra{i} x_{rq} x_{qp} \ket{p} + \sum_{i = p}^{q - 1} \eta_i \bra{i} x_{qr} x_{rp} \ket{p} 
\, .
\ee
Similar representations were found for all tree N${}^k$MHV amplitudes \cite{DruHen08}.

Analogously to the ordinary superconformal symmetry, some of the dual superconformal generators become anomalous at loop level \cite{DruHenKorSok07}. 
The explicit realization of the dual superconformal symmetry was suggested in terms of super Wilson loops \cite{ManSki10,Car10} that admit the following 
generic  form \cite{OogRahRobTan00}
\be
\vev{\mathcal{W}_n (x_i, \theta_i; a)}
= 
\frac{1}{N_c} \VEV{\tr P \exp 
\left( 
\frac{1}{2} ig \int_{\mathcal{C}_n} dx^{\dot\alpha\alpha} \mathcal{A}_{\alpha\dot\alpha} + ig \int_{\mathcal{C}_n} d \theta^{\alpha A} \mathcal{F}_{\alpha A} 
\right)}
\, ,
\ee
in terms of the bosonic $\mathcal{A}_{\alpha\dot\alpha} $ and fermionic $\mathcal{F}_{\alpha A}$ connections that are power series in the Grassmann
variables $\theta$ with $x$-dependent coefficients being the propagating fields of the $\mathcal{N} = 4$ SYM theory. The super Wilson loop is stretched 
on a contour $\mathcal{C}_n$ in superspace formed by segments connecting the vertices located at $(x_i, \theta_i)$ that are connected by straight lines. 
This is a generalization of an earlier conjecture for the dual representation for the MHV amplitudes \cite {AldMal08,KorDruSok08,BraHesTra07,KorDruHenSok08}
obtained from above by setting all Grassmann variables to zero. The former is well tested by now confronting it against multi-loop/multileg results on the 
amplitude side \cite{ABDK03, BDS05, AnaBraHesKhoSpeTra09, DelDuhSmi10, GonSprVerVol10, BerDixKosRoiSprVerVol08, CacSprVol08}.
In the latter case, the bosonic Wilson loop has a polygonal contour  $C = [x_1, x_2] \cup [x_2, x_3] 
\dots [x_n, x_1]$ with its sites defined by particle momenta $p_i = x_{ii+1}$. The dual superconformal symmetry of superamplitudes is realized as ordinary 
superconformalsymmetry of the super Wilson loop. The precise correspondence between the latter and non-MHV amplitudes is purpoted to work as follows. 
The expansion of $\mathcal{W}_n$ in terms of fermionic variables goes in powers of $\theta^4$
\be
\label{GrassmannWexp}
\vev{\mathcal{W}_n} = \vev{\mathcal{W}_{n;0}} + \vev{\mathcal{W}_{n;1}} + \dots
\, ,
\ee
with $\vev{\mathcal{W}_{n;k}}$ being a homogeneous polynomial of order $4 k$, and formally can go all the way to $\theta^{4n}$. Then the fulfillment of the
duality between the two objects requires
\be
\vev{\mathcal{W}_{n;k} (x_i, \theta_i; a)} = a^k \widehat{\mathcal{A}}_{n;k} (\lambda_i, \widetilde\lambda_i, \eta_i; a)
\, ,
\ee
and all components with $k > n - 4$ vanishing. However, while this correspondence was verified for the lowest component 
$\vev{\mathcal{W}_{n;0}}$ by explicit strong \cite {AldMal08} and weak \cite{KorDruSok08,BraHesTra07,KorDruHenSok08} coupling analyses 
that confirmed the duality for MHV amplitudes, the foundation of the above 
matching for super Wilson loop in Minkowski space with non-MHV amplitudes faced difficulties and requires further elucidation. Namely, the 
consideration performed in Ref.\ \cite{BelKorSok11} demonstrated that due to on-shell nature of the closure of the supersymmetry
algebra in maximally supersymmetric Yang-Mills theory and highly singular nature of superloop on light-like polygonal contour, the Poincar\' e 
supersymmetry gets broken already at one loop order by effects stemming from the field equations of motion when one uses a version of dimensional
regularization known as Four-Dimensional Helicity scheme \cite{BerFreDixWon02}. The same anomalous contribution 
violates conformal symmetry thus invalidating the naive duality between the superamplitude and super Wilson loop. 

However, as was already pointed out in Ref.\ \cite{BelKorSok11}, the supersymmetric Wilson loop is a scheme dependent quantity inheriting the 
property from its purely bosonic progenitor. The very fact of the scheme dependence of the super Wilson loop can be verified by applying the 
operator product expansion (OPE) reasoning to the multipoint correlation functions of the stress-tensor supermultiplet $\mathcal{T}$. These were 
identified as yet another equivalent representation of non-MHV scattering amplitudes in Refs.\ \cite{EdeHesKorSok11} and checked in trees and loops. 
Again this generalized the conjecture valid for the MHV case \cite{EdeKorSok09}. As a consequence the resolution of the above puzzle is hiding in the 
fact that only the product of the OPE coefficient function and the vacuum expectation value of the superloop is well-defined. Thus one may reshuffle the 
anomalous contributions away from the latter into the former. This is alike a finite scheme transformation. Similar phenomena are not 
new and were studied previously in other circumstances as we will discuss below. 

Since the discussion parallels the light-cone conformal OPE for two-point functions, let us discuss it first since we will acquire from it the intuition on the 
scheme-dependent pattern of conformal symmetry breaking. It suffices to limit the consideration to the leading-twist, i.e., twist two, contribution to the OPE
for the product of two dimension-$d_{\mathcal{O}}$ protected operators as $x^2 \to 0$
\be
\label{COPE}
\mathcal{O} (x) \mathcal{O} (0) 
= 
\frac{1}{(x^2)^{d_\mathcal{O}-1}} \sum_{j} C_j (\mu^2 x^2; a) \mathbb{O}_j (0; \mu^2) + \dots
\, ,
\ee
where the ellipses stand for contributions of higher twist operators. The sum in the right-hand side runs over the conformal spin $j$ of contributing 
operators of canonical dimension $d_j$. The coefficient function $C_j$ admits an infinite series expansion in 't Hooft coupling $a$ and depends on 
the dimensionless product of the near light-like distance $x^2$ and the factorization scale $\mu^2$. The twist-two conformal operators $\mathbb{O}_j$ 
develop corresponding  dependence on the renormalization scale $\mu^2$ due to their nonvanishing anomalous dimensions $\gamma_j (a)$. The 
above expansion separates the physics of short and long light-cone distances, encoding the former/latter into the coefficient function/conformal
operator. In writing above OPE, one tacitly assumes that the conformal covariance is preserved by the perturbative scheme 
adopted in the analysis of ultraviolet divergences in the renormalization of conformal operators  $\mathbb{O}_j$, such that the latter obey 
autonomous renormalization group equations
\be
\label{Nomixing}
\frac{d}{d \log \mu} \mathbb{O}_j = - \gamma_j  (a) \mathbb{O}_j 
\, ,
\ee
and as a consequence, the coefficient function can be factorized into a perturbatively corrected Clebsh-Gordan coefficient in the decomposition 
of the product of two dimension-$d_\mathcal{O}$ representation into irreducible components which is expressed in terms of the confluent
hypergeometric function \cite{FerGatGri73}
\be
C_j (\mu^2 x^2; a) = c_j (a) (\mu^2 x^2)^{\gamma_j (a)} 
{{}_1F_1} \Big(\ft12(d_j + j + \gamma_j (a) + 1), d_j + j + \gamma_j (a) + 1; x \cdot \partial \Big)
\, ,
\ee
and a coupling-dependent coefficient $c_j (a)$. 

It is important to realize that while the left-hand side of the \re{COPE} is independent of the factorization scale $\mu$ and is scheme independent,
the OPE introduces ultraviolet divergences in individual terms and thus dependence on the way they are treated perturbatively for both conformal
operators and coefficient functions. This scheme dependence is intrinsic to the anomalous dimensions of conformal operators. Namely, using 
conventional regularization procedures based on a departing from four space-time dimensions, be it the original dimensional regularization or 
dimensional reduction etc., and the (modified) minimal subtraction scheme, one immediately discovers that conformal operators mix with each 
other starting from two-loop order,
\be
\frac{d}{d \log \mu} \mathbb{O}^{\rm\scriptscriptstyle MS}_j = - \sum_{k \leq j} \gamma_{jk}  (a) \mathbb{O}_k^{\rm\scriptscriptstyle MS}
\, ,
\ee
with $\gamma_{j > k}  (a) \sim a^2$. This phenomenon was well studied in QCD \cite{BelMul98}. In this theory, the reason for mixing is two-fold. First, the 
conformal symmetry gets broken by the renormalization of the QCD strong coupling which induces non-trivial beta function, $\beta \neq 0$. Second, there 
is another effect, present in any theory, conformal or not, which is more subtle. This is the effect that we are after here being interested in application
to $\mathcal{N} = 4$ theory. Whenever dimensional regularization is used the conformal invariance is broken even if  the four-dimensional beta 
function $\beta$ vanishes since the coupling becomes dimensional away from four dimensions $g \to g_\varepsilon = g \mu^\varepsilon$ 
and therefore $\beta_\varepsilon = - \varepsilon g$. This $O (\varepsilon)$ effect when accompanied by $1/\varepsilon$ poles in perturbative graphs 
generates finite, anomalous contributions. The studies of conformal Ward identities for the correlation function with conformal operator insertion allowed 
one to compute a (finite) scheme transformation matrix $\mathcal{B}$ to the conformally covariant basis of operators obeying Eq.\ \re{Nomixing},
\be
\mathbb{O}^{\rm\scriptscriptstyle MS}_j = \sum_{k \leq j} \mathcal{B}_{jk} \mathbb{O}_k
\, .
\ee
Here the matrix $\mathcal{B}$ is determined by the special conformal anomaly matrix $\gamma^c_{jk}$ as follows $\mathcal{B}_{jk} = - a_{jk}^{-1} 
\sum_{l <  j} \gamma^c_{jl} \mathcal{B}_{lk}$ and arises from the renormalization of the product of the operator $\mathbb{O}^{\rm\scriptscriptstyle MS}_j$ 
and conformal boost variation of the regularized action
\be
\mathbb{O}^{\rm\scriptscriptstyle MS}_j ( \delta_\kappa S ) = \sum_{k \le j} \gamma^c_{jk} (a) \mathbb{O}^{\rm\scriptscriptstyle MS}_k
\, .
\ee
Thus $\mathcal{B}$ diagonalizes the mixing matrix $\gamma_{jk}$ by removing from it conformal symmetry breaking effects.

Now, let us turn to the correlation functions of the stress-tensor supermultiplet $\mathcal{T}$  \cite{EdeHesKorSok11},
\be
G_n (x_i, \theta_i; a) = \vev{\mathcal{T} (x_1, \theta_1) \dots \mathcal{T} (x_n, \theta_n)}
\, ,
\ee
with anti-chiral Grassmann variables set to zero. Recently, it demonstrated found that their multiple pairwise light-cone limit $x_{ii+1}^2 \to 0$ is related to 
the square of the full superamplitude \cite{EdeHesKorSok11}
\be
\lim_{x_{ii+1}^2 \to 0} G_n (x_i, \theta_i; a) \sim \left( \prod_{i=1}^n x_{ii+1}^{-2} \right) 
\left( \sum_{k=0}^{n-4} a^k \widehat{\mathcal{A}}_{n,k} (\lambda_i, \widetilde\lambda_i, \eta_i; a) \right)^2
\, .
\ee 
It is important to realize that the correlation function is well-defined 
in four dimensions away from the light-cone limit and thus does not require a regularization. However, by taking the limit, one effectively constructs an OPE 
similar to the previously discussed two-point case. Then the leading term in the expansion is given by the product of the (square of the) vacuum expectation 
value of the super Wilson loop and a coefficient function,
\be
\lim_{x_{ii+1}^2 \to 0} G_n (x_i, \theta_i) 
= 
\left( \prod_{i=1}^n x_{ii+1}^{-2} \right) C \big(g; \mu^2 x_{ij}^2 , \theta_i \big) \vev{ \mathcal{W}_n (x_i, \theta_i)}^2
\, .
\ee
The former, being an eikonal approximation for particle's propagation, encodes the long-distance physics. On the other hand, the latter takes
care of the rest. Therefore, while the product of the two is a scheme independent observable, each of them separately does depend on the
on the way to handle the divergences. The Wilson loop is singular even away from the light-cone limit due to the mere presence of the cusps 
on its contour be it lights like or not. This is pertinent to the purely bosonic case where the identification of the scattering amplitudes with the 
expectation value of the Wilson loop on a light-light polygonal contour requires a specific choice for identification of the infrared scale parameter 
on the amplitude side with the ultraviolet one, on the other. This scheme dependence becomes trickier as one ascends to superspace. Though 
the final result of one-loop calculations of the super Wilson loop corresponding to NMHV level of scattering amplitudes is finite, the intermediate 
steps require a regularization procedure. What this implies is that the latter breaks the conformal symmetry at the intermediate steps and this effect 
can permeate into the final answer as an anomaly. This is one-to-one correspondence with the argument alluded to above for the conformal 
operators. The difference is however that the anomaly contributes starting already from one loop which is a consequence of the more singular nature
of the light-like Wilson loop: have the one-loop renormalization of conformal operators induce a double pole in $\varepsilon$, the anomalous
dimension matrix of the latter would be off-diagonal already at one loop, not two.

The desire to use helicity formalism for scattered particles severely constraints the choice of acceptable regularization schemes. Since one 
conventionally uses a variation of dimensional regularization for calculations of scattering amplitudes which allows one, on the one hand, to 
preserve supersymmetry and, on the other, straightforward generalization beyond one-loop order, it is paramount to adopt the same scheme for 
perturbative analysis of dual observables. As we advocated earlier, the use of a version of dimensional regularization, known as FDH scheme, 
which yields anomalous contributions in Wilson loop expectation values thus spoiling the duality between them and the non-MHV scattering 
amplitudes. Therefore, in order to restore it, one has to subtracting the anomalous terms computing the latter from conformal Ward identities.
By an explicit calculation, we will demonstrate below that at one loop in the Four-Dimensional Helicity scheme
\be
\label{NMHVandAnomaly}
a \, \widehat{\mathcal{A}}_{n;1} = \vev{\mathcal{W}_{n;1}}  - a_{n;1}
\, ,
\ee
where $a_{n;1}$ is the anomalous contribution arising from breaking of the special conformal boost by the regularization procedure at loop level. Notice 
that due to the specific origin of ultraviolet divergences in the super Wilson loop as arising from the light-like nature of the polygon sides connecting 
nearest pairs of vertices, only the graphs were the virtual particles are exchanged (at one loop) between the nearest and next-to-nearest links require
regularization. All other contributions, involving exchanges separated by more than one site, are finite. This implies that since each link carries 
two Grassmann variables associated with them, the anomalies will be present in only for NMHV amplitudes with four adjacent Grassmann variables. 

Our subsequent presentation is organized as follows. In the next section, we remind the definition of the super Wilson loop along with conventions
used. In Section \ref{OneLoopSuperW}, we perform one-loop computations of the $\vev{\mathcal{W}_{n,1}}$ component of the super Wilson loop
for $n=5,6$ to support the structure suggested by Eq.\ \re{NMHVandAnomaly}. Then in Section \ref{ConfWI}, we construct conformal Ward identities for
the super Wilson loop and calculate one-loop conformal anomaly for all tree-level NMHV amplitudes. We demonstrate that indeed it explains the 
anomalous nature of the loop and necessitates a finite subtraction to restore conformal symmetry. Finally, we comment on our results. 

\section{Superymmetric Wilson loop at one loop}
\label{OneLoopSuperW}

The supersymmetric Wilson loop that we presently study is defined by the path-ordered product of the links $\mathcal{W}_{[ii+1]}$ 
\be
\vev{\mathcal{W}_n} = \frac{1}{N_c} \vev{\tr \left( \mathcal{W}_{[12]} \dots \mathcal{W}_{[n1]} \right)}
\ee
on segments of  the polygonal contour $\mathcal{C}_n$ that are parametrized as
\be
x_{[i i+1]} = x_i - t x_{i i+1} \, , \qquad \theta_{[i i+1]} = \theta_i - t \theta_{i i+1}
\, .
\ee
The individual superlines are exponents of the line integrals of superconnections
\be
\mathcal{W}_{[i i+1]} = P {\rm e}^{i g E_{[i i+1]}}
\, ,
\ee
with $E_{[i i+1]}$ admitting an expansion in Grassmann variables
\be
\label{SuperConnectionExpansion}
E_{[i i+1]} = \sum_n E_{[i i+1]}^{[n]}
\, ,
\ee
with each term in the sum $E_{[i i+1]}^{[n]}$ being a polynomial of order $n$ in $\theta$'s. They read explicitly (up to the fourth order)
\ba
E_{[i i+1]}^{[0]} 
&=& 
- \ft12 \int_0^1 dt \, \bra{i} A |i]
\, , \nonumber\\
E_{[i i+1]}^{[1]} 
&=& 
- \ft{i}{2} \chi_i^A \int_0^1 dt \, [ \bar\psi_A \, i]
\, , \nonumber\\
E_{[i i+1]}^{[2]} 
&=& 
- \ft{i}{2} \chi_i^A \int_0^1 dt \, \Big( \ft12 \bra{\theta^B_{[i i+1]}} D |i]+ \eta_i^B \Big) \bar\phi_{AB}
\, , \nonumber\\
E_{[i i+1]}^{[3]} 
&=& 
- \ft13 \varepsilon_{ABCD} \chi_i^A \int_0^1 dt \, 
\Big( \ft14 \bra{\theta^B_{[i i+1]}} D |i] + \eta_i^B \Big) \vev{\theta^C_{[i i+1]} \psi^D}
\, , \nonumber\\
E_{[i i+1]}^{[4]} 
&=& 
- \ft{i}{8} \varepsilon_{ABCD} \chi_i^A \int_0^1 dt \, 
\Big( \ft16 \bra{\theta^B_{[i i+1]}} D |i] + \eta_i^B \Big) \bra{\theta^C_{[i i+1]}} F \ket{\theta^D_{[i i+1]}}
\, .
\label{connection}
\ea
Here, we introduced new notations so a few comments are in order. First, we wrote the above relations in terms of the fermionic components 
$\chi_i^A$ of momentum supertwistors \cite{Hod09} related to the Grassmann coordinates of the contour as $\chi_i^A = \vev{i \theta_i^A}$.
We use the bra and ket formalism to write down the inner products, such that by defining $\bra{A} = A^\alpha$, $\ket{A}_\alpha$, $[A| = A_{\dot\alpha}$
and $|A] = A^{\dot\alpha}$ we have $\vev{AB} = A^\alpha B_\alpha$ and $[AB] = A_{\dot\alpha} B^{\dot\alpha}$. Then inverting the relation for 
$\chi_i^A$ in favor of fermionic coordinates of the cusps $\theta_{i \alpha}^A$, one finds
\be
\ket{\theta_i^A} = \frac{\chi_{i-1}^A \ket{i} - \chi_i^A \ket{i-1}}{\vev{i-1 i}}
\, .
\ee
This immediately provides a relation of $\chi$'s to the Grassmann variables on the superamplitude side,
\be
\label{ChiToEta}
\ket{\theta_{i i+i}^A} = \eta_i^A \ket{i} 
\, , \qquad
\eta_i^A = \frac{\chi_{i-1}^A}{\vev{i-1 i}} + \frac{\chi_{i+1}^A}{\vev{i i+1}} - \frac{\vev{i-1 i+1}}{\vev{i-1 i}\vev{i i+1}} \chi_i^A
\, .
\ee
Next, each term in the line integral over the superconnections goes over the fields populating the $\mathcal{N}=4$ SYM supermuliplet,
the gauge field $A_{\alpha\dot\alpha}$, the gaugino $\bar\psi_{\dot\alpha A}$, the scalars $\phi_{AB}$ in the {\bf 6} of SU(4), the conjugate 
gaugino $\psi^{\alpha A}$ and the chiral component of the gluon field-strength tensor $F^{\alpha\beta} = \ft14 (\sigma_{\mu\nu})^{\alpha\beta} F^{\mu\nu}$.
The covariant derivatives $D_{\alpha\dot\alpha} = \partial_{\alpha\dot\alpha} - i g [A_{\alpha\dot\alpha}, \, ]$ in the above equations act on the argument 
$x_{[ii+1]}$ of the fields only. Finally, the subleading terms that we do not display in the expansion in Eq.\ \re{SuperConnectionExpansion}, which goes up 
to $\theta^8$, are either interaction-dependent and thus vanish in the free-field limit or proportional to the equation of motion. Since in the present paper 
we will be after the $\mathcal{W}_{n;1}$ term, none of these will be of relevance for our analysis.

The one-loop calculation of super Wilson loop was already performed in Ref.\ \cite{BelKorSok11} for four points. Contrary to the expectation
based on the conjectured duality between the amplitudes and supersymmetric Wilson loop, the found result was nonvanishing. That calculation was done
making use the standard FDH regularization scheme especially well adopted for helicity formalism used in handling gauge theory
amplitudes. The discovered result was purely anomalous. It was simultaneously breaking Poincar\'e supersymmetry and special conformal boosts. However,
the consistency of the regularization procedure was demonstrated by derivation of supersymmetric Ward identities and an independent calculation of
arising supersymmetric anomalies. Below, extending earlier considerations \cite{BelKorSok11}, we will demonstrate that the anomalous contribution to
super Wilson loop can be isolated and subtracted out in the fashion that we advocated in the Introduction.

In order to understand the structure of the result, without obscuring it with unnecessary details, we will focus in this section only on the $\chi_1^4$ contribution%
\footnote{Here and everywhere in the paper, we introduce a shorthand notation $\chi_i \chi_j \chi_k \chi_l \equiv \varepsilon_{ABCD} \chi_i^A \chi_j^B 
\chi_k^C \chi_l^D$ to simplify presentation of formulas.}  to the supersymmetric Wilson loop expectation value. The conjectured duality claims that it should 
be equal to tree gluon NMHV amplitude multiplied by the factor of the 't Hooft couping, more precisely to $\chi_1^4$ component of the sum of $R$-superinvariant 
defining it at tree level \re{treeNMHV}. Extracting the component in question from the result of our earlier work \cite{BelKorSok11}, we find
\be
\vev{\mathcal{W}_{4;1}} = - \frac{a}{48} \frac{[42]}{[41][12]} \frac{\vev{42}^3}{\vev{41}^3 \vev{12}^3} \chi_1^4 + \dots
\, .
\ee
Below we compute one-loop contribution to the same component for the pentagon and hexagon and determine the common pattern resulting from it. As a 
consequence we will be able to generalize our consideration to any number of points.

\subsection{Pentagon}

\begin{figure}[t]
\begin{center}
\mbox{
\begin{picture}(0,145)(190,0)
\put(0,25){\insertfig{13}{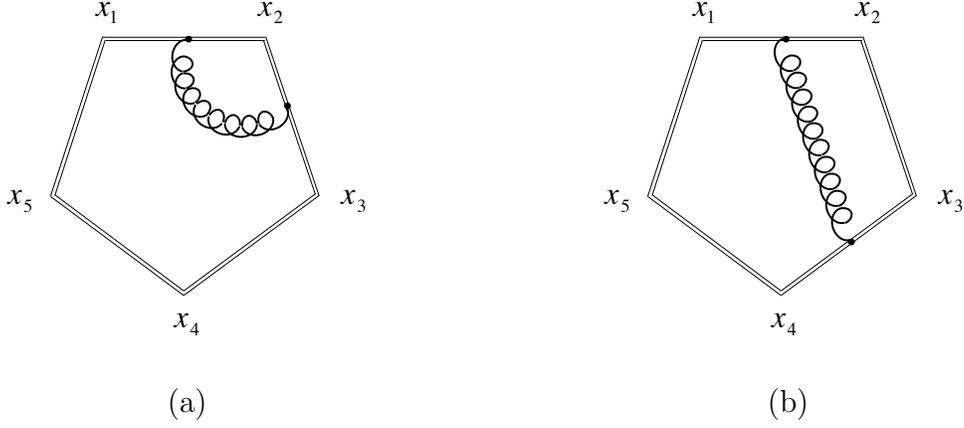}}
\put(65,0){(a)}
\put(292,0){(b)}
\end{picture}
}
\end{center}
\caption{ \label{Pentagon} One-loop contribution to $\chi_1^4$ component of the pentagon supersymmetric Wilson loop.}
\end{figure}

First, let us begin with the pentagon. At one-loop order, the contribution to the component in question arises from just two components of the 
superconnections, such that the expectation value is given by the following correlation function
\be
\vev{\mathcal{W}_{5;1}} = \frac{(ig)^2}{N_c} \sum_i \vev{\tr \big( E_{[12]}^{[4]} E_{[i i+1]}^{[0]} \big)}
\, ,
\ee
where $E^{[0]}$ and $E^{[4]}$ are determined by the first and last equation in \re{connection}, respectively. This function receives nontrivial contributions 
from the vertex and exchange diagrams, displayed in Fig.\ \ref{Pentagon} (a) and (b), respectively, and their mirror reflections $(\bar{\rm a})$ and 
$(\bar{\rm b})$. As we already advertised in the Introduction, in order to perform the analysis efficiently we will use the rules of the FDH regularization 
scheme \cite{BerFreDixWon02}. Its main advantage is that while being a version of dimensional regularization, it allows one to perform the spinor 
decomposition of supermomenta even in regularized theory and rely on four-dimensional manipulation rules like Fierz transformation etc. The propagator 
entering the computation reads
\baa
\vev{A_{\alpha\dot\alpha} (x_1) F_{\gamma\delta} (x_2)}
=
i \frac{\Gamma (2 - \varepsilon)}{4 \pi^{2 - \varepsilon}}
\frac{1}{[-x_{12}^2]^{2 - \varepsilon}}
(
\varepsilon_{\alpha\delta} (x_{12})_{\dot\alpha\beta}
+
\varepsilon_{\alpha\beta} (x_{12})_{\dot\alpha\delta}
)
\, .
\eaa
Notice that individual graphs induce logarithmic terms depending on Mandelstam invariants which cancel only in the sum of \ref{Pentagon} 
(b) and its mirror $(\bar{\rm b})$. We will not display them in intermediate formulas that follow. The individual contributions read respectively for 
the graphs in Fig.\ \ref{Pentagon} (a) and its mirror
\ba
\vev{\mathcal{W}_{5;1}^{({\rm a})+(\bar{\rm a})}}
\!\!\!&=&\!\!\!
- \frac{a}{48} \frac{\vev{52}^2 \chi_1^4}{\vev{51}^2 \vev{12}^2} \left\{ \frac{1}{x_{13}^2} + \frac{1}{x_{52}^2} \right\}
\, , 
\ea
as well as diagrams (b) and its mirror $(\bar{\rm b})$
\ba
\vev{\mathcal{W}_{5;1}^{({\rm b})}}
\!\!\!&=&\!\!\!
- \frac{a}{48} \frac{\chi_1^4}{\vev{51}^3 \vev{12}^2}
\frac{\bra{5} x_{53} |3]}{x_{13}^2 (x_{14}^2 - x_{24}^2)}
\bigg\{
\vev{13} \vev{52}^2
-
\vev{23} \bra{5} x_{63} |3]
\left(
\frac{\vev{12} \vev{35}}{x_{14}^2} - \frac{\vev{23} \vev{51}}{x_{24}^2}
\right)
\bigg\}
\, , \nonumber\\
&&\\[-1mm]
\vev{\mathcal{W}_{5;1}^{(\bar{\rm b})}}
\!\!\!&=&\!\!\!
- \frac{a}{48} \frac{\chi_1^4}{\vev{51}^2 \vev{12}^3}
\frac{\bra{2} x_{42} |4]}{x_{52}^2 (x_{42}^2 - x_{41}^2)}
\bigg\{
\vev{41} \vev{52}^2
-
\vev{45} \bra{2} x_{42} |4]
\left(
\frac{\vev{51} \vev{24}}{x_{42}^2} - \frac{\vev{12} \vev{45}}{x_{41}^2} 
\right)
\bigg\}
\, . \nonumber
\ea
Making use of the identity 
$$
\vev{12} \vev{13} \frac{\bra{5} x_{53} |3]}{x_{13}^2} - \vev{51} \vev{41} \frac{\bra{2} x_{42} |4]}{x_{24}^2} = - (x_{14}^2 - x_{24}^2) \frac{x_{35}^2}{[51] [12]}
\, ,
$$ 
it allows one to rewrite the sum of the last two contributions in a form free from the spurious poles at $x_{14}^2 = x_{24}^2$, such that it can be cast in
the form
\ba
\vev{\mathcal{W}_{5;1}^{({\rm b})+(\bar{\rm b})}}
&=&
\frac{a}{48} \frac{\vev{52}^2 \chi_1^4}{\vev{51}^2 \vev{12}^2} \frac{x_{35}^2}{x_{13}^2 x_{52}^2} 
\left\{ 
1+
\frac{[34]^2}{x_{24}^2 x_{41}^2} 
\frac{\vev{23}^2 \vev{45}^2}{\vev{52}^2}
\right\} 
\, .
\ea
Adding all these terms together and using the identity $x_{35}^2 = x_{13}^2 + x_{52}^2 - \vev{52} [52]$, one finds that the pentagon Wilson loop reads
\be
\label{W5total}
\vev{\mathcal{W}_{5;1}} = - \frac{a}{48} \chi_1^4
\left\{ 
\frac{[52]}{[51][12]} \frac{\vev{52}^3}{\vev{51}^3 \vev{12}^3}
-
\frac{x_{35}^2 [34]^2}{x_{13}^2 x_{24}^2 x_{41}^2 x_{52}^2}
\frac{\vev{23}^2 \vev{45}^2}{\vev{51}^2 \vev{12}^2}
\right\}
\, .
\ee
A naked eye inspection immediately reveals that the first term in the curly brackets is identical (making use of the obvious replacement $4 \to 5$ since 
now the point $5$ becomes adjacent to $1$) to the anomalous contribution defining the four-cusp Wilson loop quoted in the preamble to this section. Then 
the remaining term has to be identical to the five-point reduced NMHV amplitude $\widehat{\mathcal{A}}_{5;1}$. Indeed, for five points the
NMHV amplitude is nothing else but the $\overline{\rm MHV}$, i.e, conjugate to MHV. However, for subsequent generalization to more points it is 
instructive to recall its expression in terms of superconformal invariants. Namely, the tree five-particles NMHV amplitude is simply determined by
one $R$-invariant and reads \cite{DrunHenKorSok08b}
\be
\widehat{\mathcal{A}}_{5,1} = R_{241}
\, .
\ee
The complicated general form of the latter simplifies enormously for five points and takes the following well-known form 
\be
R_{241} = \delta^{(4)} (\eta_1 [23] + \eta_2 [13] + \eta_3 [21]) 
\frac{\vev{12} \vev{23} \vev{34} \vev{45} \vev{51}}{\vev{45}^4 [12][23][34][45][51]} 
\, .
\ee
Cyclic invariance of the amplitude implies that $R_{241} = R_{413}$ upon the use of the (super)momentum conserving delta-functions, which provides an 
alternative but identical representation. Expanding the $R$-invariant in components one finds for the contribution in question
\be
R_{241} = - \frac{x_{35}^2 [34]^2}{x_{13}^2 x_{24}^2 x_{41}^2 x_{52}^2}\frac{\vev{23}^2 \vev{45}^2}{\vev{51}^2 \vev{12}^2} \chi_1^4 + \dots
\, .
\ee
It is thus readily identified with the second term in the curly brackets of Eq.\ \re{W5total}.

\subsection{Hexagon and up}

\begin{figure}[t]
\begin{center}
\mbox{
\begin{picture}(0,145)(250,0)
\put(0,25){\insertfig{17}{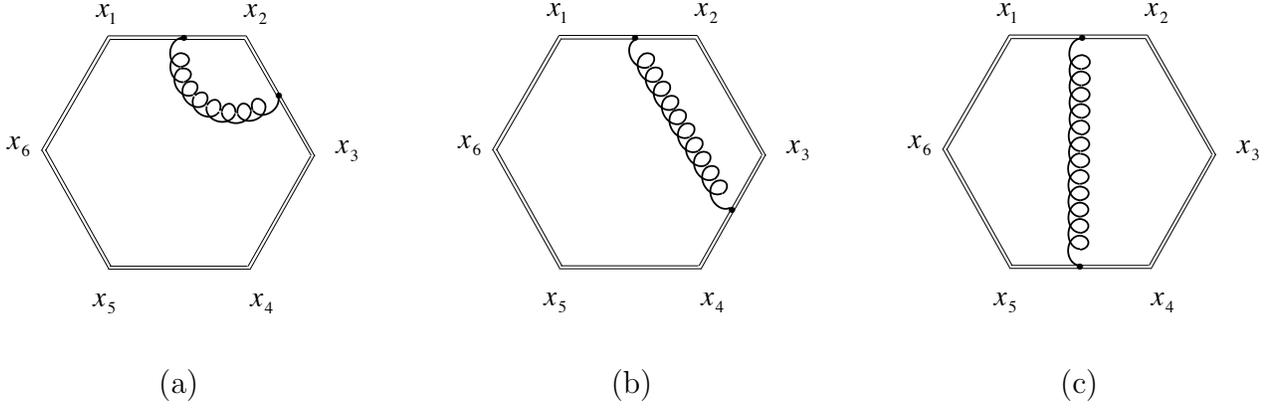}}
\put(62,0){(a)}
\put(233,0){(b)}
\put(403,0){(c)}
\end{picture}
}
\end{center}
\caption{ \label{Hexagon} Non-vanishing one-loop $\chi_1^4$-contribution to the hexagon Wilson loop.}
\end{figure}

Now we address a less trivial example which involves a genuine NMHV amplitude, the hexagon. The nonvanishing Feynman diagrams that
define the component in question are shown in Fig.\ \ref{Hexagon}. As for the pentagon case, we will display only rational contribution from 
each of the graph, ignoring the logarithmic terms which cancel in the total sum. The calculations are straightforward for Fig. \ref{Hexagon} (a) 
and its mirror and yield
\ba
\vev{\mathcal{W}_{6;1}^{({\rm a})+(\bar{\rm a})} }
&=&
- \frac{a}{48} \frac{\chi_1^4 \vev{62}^2}{\vev{61}^2 \vev{12}^2}
\bigg\{
\frac{1}{x_{13}^2} + \frac{1}{x_{62}^2}
\bigg\}
\, ,
\ea
as well as for Fig. \ref{Hexagon} (b) and its mirror, respectively, one finds
\ba
\vev{\mathcal{W}_{6;1}^{({\rm b})}}
\!\!\!&=&\!\!\!
- \frac{a}{48} \frac{\chi_1^4}{\vev{61}^3 \vev{12}^2}
\frac{\bra{6} x_{63} |3]}{x_{13}^2 (x_{14}^2 - x_{24}^2)}
\bigg\{
\vev{13} \vev{62}^2
-
\vev{23} \bra{6} x_{63} |3]
\left(
\frac{\vev{12} \vev{36}}{x_{14}^2} - \frac{\vev{23} \vev{61}}{x_{24}^2}
\right)
\bigg\}
\, , \nonumber\\
&&\\[-1mm]
\vev{\mathcal{W}_{6;1}^{(\bar{\rm b})}}
\!\!\!&=&\!\!\!
- \frac{a}{48} \frac{\chi_1^4}{\vev{61}^2 \vev{12}^3}
\frac{\bra{2} x_{52} |5]}{x_{62}^2 (x_{52}^2 - x_{51}^2)}
\bigg\{
\vev{51} \vev{62}^2
-
\vev{56} \bra{2} x_{52} |5]
\left(
\frac{\vev{61} \vev{25}}{x_{52}^2} - \frac{\vev{12} \vev{56}}{x_{51}^2} 
\right)
\bigg\}
\, . \nonumber
\ea
The exchange diagram in Fig. \ref{Hexagon} (c) require a more tedious analysis and multiple use of identities for angle and square brackets to
finally bring the result to a concise form
\ba
\vev{\mathcal{W}_{6;1}^{({\rm c})}}
&=&
\frac{a}{48} \frac{\chi_1^4}{\vev{61}^3 \vev{12}^3}
\bigg\{
\vev{14} \vev{62}^2 \frac{\vev{23} \vev{61} [34] - \vev{12} \vev{56} [45]}{(x_{14}^2 - x_{24}^2) (x_{52}^2 - x_{51}^2)}
\\
&&\qquad\qquad\qquad\quad
+
\frac{\bra{1} x_{14} |4]}{x_{14}^2 x_{52}^2 - x_{24}^2 x_{51}^2}
\bigg[
\frac{\vev{23}^2 \bra{6} x_{63} |3]^2}{x_{14}^2 - x_{25}^2}
\left(
\frac{\vev{12} \vev{46}}{x_{14}^2} - \frac{\vev{61} \vev{24}}{x_{24}^2}
\right)
\nonumber\\
&&\qquad\qquad\qquad\qquad\qquad\qquad\qquad\ 
+
\frac{\vev{56}^2 \bra{2} x_{52} |5]^2}{x_{52}^2 - x_{51}^2}
\left(
\frac{\vev{12} \vev{46}}{x_{52}^2} - \frac{\vev{61} \vev{24}}{x_{51}^2}
\right)
\bigg]
\bigg\}
\, . \nonumber
\ea
Notice that for uniformity of presentation, the numerator in the first line can be further rewritten in the form mimicking the rest of the result 
$\vev{23} \vev{61} [34] - \vev{12} \vev{56} [45] = \vev{62} [4| x_{41} \ket{1} + \vev{61} \vev{12} [14]$ and use the identity $\vev{14} [14] = 
x_{14}^2 - x_{24}^2 - x_{51}^2 + x_{52}^2$. While the denominator in second line is a ``square'' $x_{14}^2 x_{52}^2 - x_{24}^2 x_{51}^2 
= \bra{1} x_{14} |4] \bra{4} x_{14} |1]$. After a lengthy calculation, one finds for the sum of all terms
\ba
\vev{\mathcal{W}_{6;1}}
&=& 
- \frac{a}{48} \chi_1^4
\bigg\{
\frac{[62]}{[61][12]} \frac{\vev{62}^3}{\vev{61}^3 \vev{12}^3}
\\
&+&
\frac{\vev{23} \vev{34} \vev{45} \vev{56}}{\vev{61}^3 \vev{12}^3 \bra{4} x_{14} |1]}
\bigg[
\frac{\bra{6} x_{63} |3]^3}{\vev{45} \vev{56} [12] [23] x_{14}^2}
-
\frac{\bra{6} x_{52} |3]^3}{\vev{23} \vev{34} [56] [61] x_{52}^2}
\bigg]
\bigg\}
\, . \nonumber
\ea
Again, the first term is the anomalous contribution, which has the same universal form as for the four- and five-point Wilson loops, while the second 
term is a known two-term representation of the NMHV amplitude \cite{BriFenSprVol05} obtained by solving the BCFW recursion relations 
\cite{BriCacFenWit04}. It can be re-written in terms of the $\chi_1^4$ component of the sum of the $R$-invariants defining the six-particle 
NMHV amplitude $R_{146} + R_{136} + R_{135}$ \cite{DruHenKorSok07}. The latter equivalent to the three-term form of the gluon scattering 
amplitude originally computed in Ref.\ \cite{ManParXu88}.

As a consequence of this analysis, one can immediately generalize the consideration to more points without the need to go through explicit computations. 
Namely, the $\chi_1^4$ component of the loop Wilson loop is given by a sum of the universal anomalous contribution and the corresponding 
Grassmann component of the $R$-invariant defining the NMHV scattering amplitude,
\be
\vev{\mathcal{W}_{n;1}} 
= 
- \frac{a}{48} \chi_1^4
\bigg\{
\frac{[n2]}{[n1][12]} \frac{\vev{n2}^3}{\vev{n1}^3 \vev{12}^3}
+
\sum_{1 < q < r < n} R_{nqr} |_{\chi_1^4}
\bigg\}
\, .
\ee
Let us now turn to understanding the source of the anomalous term.

\section{Conformal anomaly}
\label{ConfWI}

In the previous sections, we observed by explicit calculations that the $\chi_1^4$ component of the super Wilson loop computed making use of the
FDH scheme develops universal anomalous contributions. The variation of this component of the super Wilson loop can be computed explicitly using 
the conformal variation of its components\footnote{Their explicit transformation
properties are $\delta_\kappa \chi_i = 0$, $\delta_\kappa (x_{ij}^2)^{\mp 1} = \pm 2 (x_{ij}^2)^{- 1} \kappa \cdot (x_i + x_j)$, $\delta_\kappa \vev{ij}^{\pm 1} 
= \pm \vev{ij}^{- 1} \bra{i} \kappa x_j + x_i \kappa \ket{j}$, $\delta_\kappa \vev{i i+1}^{\pm 1} = \pm 2 \vev{i i+1}^{\pm 1}  \kappa \cdot x_{i +1}$.},  
and reads 
\be
\label{Chi4ConfAnom}
\delta \vev{\mathcal{W}_{n;1}}
=
- \frac{a}{48} \chi_1^4 \frac{\vev{n2}^2}{\vev{n1}^2 \vev{12}^2}
\left[
\frac{1}{x_{n2}^2}
\left(
3 \bra{n} \kappa |n] - \bra{1} \kappa |n] \frac{\vev{n2}}{\vev{12}}
\right)
-
\frac{1}{x_{13}^2}
\left(
3 \bra{2} \kappa |2] - \bra{1} \kappa |2] \frac{\vev{n2}}{\vev{n1}}
\right)
\right]
\, .
\ee
In the present section we will derive this result along with all other Grassmann degree 4 contributions to the conformal anomaly of the supersymmetric
Wilson loop. To this end we derive conformal Ward identities for the latter following the approach developed in Refs.\ \cite{BelMul98} and recently 
used for the bosonic Wilson loop \cite{DruHenKorSok07} in order to fix the conformal symmetry-breaking part of the MHV amplitudes.

Writing the expectation value of superloop as a path integral
\be
\vev{\mathcal{W}_n} = \int [DX] \mathcal{W}_n {\rm e}^{i S[X]}
\, ,
\ee
with the integration performed over all fields propagating fields and ghosts, cumulatively called $X$. Then using the invariance of the path integrals under 
the conformally transformed field variables, one can easily derive the Ward identity
\be
\label{confWI}
\delta_\kappa \vev{\mathcal{W}_n} = \vev{\mathcal{W}_n (i \delta_\kappa S)}
\, ,
\ee
where vacuum average on the right-had side involves the non-vanishing conformal variation of the regularized action. The latter is of the form
\be
(i \delta_\kappa S) = - 4 i \varepsilon \int d^D z \, ( \kappa \cdot z ) \, \Delta (z) + \dots
\ee
with
\be
\label{DeltaInsertion}
\Delta (z) =  \ft12 \left[ F^{\alpha\beta} F_{\alpha\beta} + \bar{F}_{\dot\alpha\dot\beta} \bar{F}^{\dot\alpha\dot\beta} \right]
+
\ft{i}{2} 
\left[ 
\bar\psi_{\dot\alpha} ( D^{\dot\alpha\alpha} \psi_\alpha )
- 
( D_{\alpha\dot\alpha} \bar\psi^{\dot\alpha}) \psi^\alpha 
\right]
+
\ft14 ( D_\mu \phi^{AB} ) (D^\mu \bar\phi_{AB})
\, ,
\ee
where the ellipses stand for neglected BRST exact operators which do not contribute to the gauge-invariant correlation functions that we are currently 
investigating. Notice also that we ignored in the right-hand side of Eq.\ \re{confWI} the vanishing term involving the conformal boost variation of the super 
Wilson loop $\delta_\kappa \mathcal{W}_n = 0$.

The right-hand side of the Ward identity contains the anomalous term involving the conformal variation of the action which is expressed in terms 
of integrated operator insertion $\Delta (z)$ with the coordinate weight $z$. In order to extract the contribution in question is proves convenient to 
introduce the Fourier transform of the resulting correlation function 
\be
\widetilde{\mathcal{W}}_n (k) = \int d^D z \, {\rm e}^{i k \cdot z} \VEV{\mathcal{W}_n \Delta (z)}
\, ,
\ee
and while calculating it we keep only terms linear in the momentum $k$, such that the anomaly on the right-hand side of the Ward identity coms from
its second terms in the Taylor expansion
\be
\vev{\mathcal{W}_n (i \delta_\kappa S)} = - 4 \varepsilon ( \kappa \cdot \partial_k)_{k= 0} \widetilde{\mathcal{W}}_n (k)
\, .
\ee
Obviously, the anomaly admits the expansion in Grassmann variables 
\be
\widetilde{\mathcal{W}}_n = \widetilde{\mathcal{W}}_{n;0} + \widetilde{\mathcal{W}}_{n;1} + \dots
\, ,
\ee
identical to the one for the superloop itself \re{GrassmannWexp}.

\begin{figure}[t]
\begin{center}
\mbox{
\begin{picture}(0,150)(140,0)
\put(0,25){\insertfig{10}{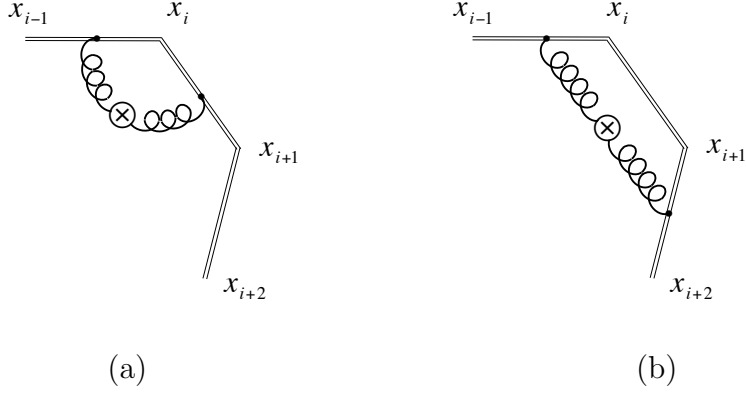}}
\put(40,0){(a)}
\put(240,0){(b)}
\end{picture}
}
\end{center}
\caption{ \label{ConfAnomaly} Feynman diagrams contributing to the conformal anomaly from gauge fields. The $\otimes$ stands for the insertion 
of the conformal variation of the action.}
\end{figure}


\subsection{Bosonic Wilson loop revisited}
\label{BosonicAnomaly}

Let us start our analysis of the one-loop calculation by revisiting the purely bosonic case \cite{DruHenKorSok07}. At one-loop, the anomaly 
$\widetilde{\mathcal{W}}_{n;0}$
\be
\label{PairwiseDecomp}
\widetilde{\mathcal{W}}_{n;0} = \sum_{i < j}^n A_{[ii+1][jj+1];0}
\ee
is given by the sum of the gluon exchanges between different sites
\be
A_{[i i+1][j j+1];0} = \frac{(i g)^2}{N_c} \int d^D z \, {\rm e}^{i k \cdot z} \VEV{ \tr E_{[i i+1]}^{[0]} E_{[j j+1]}^{[0]} \Delta^{\rm g} (z)}
\ , 
\ee
where $\Delta^{\rm g}$ is given by the first term in Eq.\ \re{DeltaInsertion} which is just the covariant gauge boson action $- \ft14 F_{\mu\nu}^2$. This 
reads in the single-gluon exchange approximation
\ba
\label{GenDelta}
A_{[i i+1][j j+1];0} 
\!\!\!&=&\!\!\! - \frac{g^2 N_c}{2} \int_0^1 ds \,  x_{i i+1}^\mu \int_0^1 dt \,  x_{jj+1}^\nu
\\
&\times&\!\!\!
\int  \frac{d^D k_1 d^D k_2}{(2 \pi)^D} \delta^{(D)} (k_1 + k_2 + k)
\frac{{\rm e}^{- i k_1 \cdot x_{[i  i+1]}}}{k_1^2}
\frac{{\rm e}^{- i k_2 \cdot x_{[j  j+1]}}}{k_2^2}
\left(
k_{1 \nu} k_{2 \mu} - g_{\mu\nu} \, k_1 \cdot k_2
\right)
\, . \nonumber
\ea
Here we used the propagator in the Feynman gauge
\be
\VEV{A_{\alpha\dot\alpha} (x_1) A_{\beta\dot\beta} (x_2)} 
=
2 i \int \frac{d^D k}{(2 \pi)^D} \frac{{\rm e}^{- i k \cdot x_{12}}}{k^2} \varepsilon_{\alpha\beta} \varepsilon_{\dot\alpha\dot\beta}
\, .
\ee
In what follows, we will be extracting terms linear in the momentum $k$ only and thus all equation below should be understood modulo
all other (irrelevant for us) contributions. Let us consider the interaction exchanges between the site $[12]$ and the rest. First, a simple 
calculation immediately yields for the $1/\varepsilon$-pole contribution arising from nearest-neighbor interaction displayed in Fig.\ \ref{ConfAnomaly} (a),
\be
\label{NearestNeighbor}
A_{[12][23];0}= - a \frac{\Gamma (1-\varepsilon)}{32\varepsilon} (- \mu^2 x_{13}^2 \pi)^\varepsilon
\left\{ 3 \bra{1} k  |1] - 3 \bra{2} k |2] - 4 k \cdot x_2 + \frac{8}{\varepsilon} k \cdot x_2 \right\}
\, .
\ee
One observes the presence of translationally non-invariant contribution in the single-pole term. The latter has to cancel against additional
Feynman diagrams. A quick inspection shows that the only other graph that also develops divergent contribution is the one where the gluon is exchanged 
between the next-to-nearest neighbor sites, see Fig.\ \ref{ConfAnomaly} (b). This implies that all $A_{[12] [k k+1];0}|_{3 < k < n-1}$ are finite. One
obtains after a straightforward calculation
\be
A_{[12][34];0}= - a \frac{\Gamma (1-\varepsilon)}{16 \varepsilon} (- \mu^2 x_{13}^2 \pi)^\varepsilon \, 
k \cdot (x_2 + x_3)
\, ,
\ee
which is not translationally invariant either. Adding the two, one finds 
\be
A_{[12][23];0} + A_{[12][34];0}
= 
- a \frac{\Gamma (1-\varepsilon)}{32 \varepsilon} (- \mu^2 x_{13}^2 \pi)^\varepsilon
\left\{ 3 \bra{1} k  |1] - 4 \bra{2} k |2]  + \frac{8}{\varepsilon} k \cdot x_2 \right\}
\, ,
\ee
where the translational invariance of the single-pole contribution is restored. On the other hand, it appears that this finding is different from the result of 
Ref.\ \cite{DruHenKorSok07}.  The resolution of this puzzle merely lies in a different arrangement of terms in the single-pole contribution. Namely,
adding mirror symmetric diagrams (with the symmetry axis going through the link-$[12]$) to the two graphs computed above, one can re-arrange terms in the
sum of next-to-nearest neighbor graphs as
\ba
A_{[12][34];0} +A_{[12][n-1 n];0} &=& - \frac{a}{16 \varepsilon} \left[ k \cdot (x_1 + x_3) + k \cdot (x_n + x_2) \right] \\
&\equiv& \bar{A}_{[12][34];0} +\bar{A}_{[12][n-1 n];0}
\, , \nonumber
\ea
such that the sum of the first term in the square brackets with Eq.\ \re{NearestNeighbor} gives the well-known form for the two-site conformal
anomaly,
\be
\label{SiteConfAnomaly}
A_{[12][23];0} + \bar{A}_{[12][34];0} 
= - a \frac{\Gamma (1-\varepsilon)}{32 \varepsilon} (- \mu^2 x_{13}^2 \pi)^\varepsilon
\left\{ 4 \bra{1} k  |1] - 4 \bra{2} k |2] + \frac{8}{\varepsilon} k \cdot x_2 \right\}
\, .
\ee
It is interesting to note that the diagram Fig.\ \ref{ConfAnomaly} (b) does not contribute to the calculation of Ref.\ \cite{DruHenKorSok07}. This is
achieved by means of a gauge transformation of the Wilson-loop links, such that each site acquires a total-derivative contribution. So effectively the
authors of Ref.\ \cite{DruHenKorSok07} do not use the Feynman gauge. Namely, the addition of total-derivative terms amounts to replacement of the 
integrand in Eq.\ \re{GenDelta} as follows,
\be
\label{SubReshuf}
k_{1 \nu} k_{2 \mu} - g_{\mu\nu} \, k_1 \cdot k_2
\ 
\to
\ 
k_{1 \nu} k_{2 \mu} - k_{1 \mu} k_{2 \nu} - g_{\mu\nu} \, k_1 \cdot k_2
\, ,
\ee
i.e., $A_{[i i+1][j j+1];0} \to A_{[i i+1][j j+1];0} - G_{[i i+1][j j+1]}$. Note that the extra contribution $k_{1 \mu} k_{2 \nu}$ produces vanishing effect in a gauge-invariant 
Wilson loop. However, what it achieves locally is to shift the effect of next-to-nearest graphs to the vertex-type diagrams in Fig.\ \ref{ConfAnomaly} (a). The reason 
for this is that all graphs but the vertex-like become finite after the substitution \re{SubReshuf} as can be seen from the explicit form of the extra term
\ba
G_{[i i+1][j j+1]} 
&=& 
- a \frac{\Gamma (1- \varepsilon)}{16 \varepsilon}
\bigg\{
(- \mu^2 x_{i+1 j+1}^2)^\varepsilon \, k \cdot (x_{i+1} + x_{j+1})
+
(- \mu^2 x_{i j}^2)^\varepsilon \, k \cdot (x_{i} + x_{j})
\nonumber\\
&&\qquad\qquad\ \ 
-
(- \mu^2 x_{i+1 j}^2)^\varepsilon \, k \cdot (x_{i+1} + x_{j})
-
(- \mu^2 x_{i j+1}^2)^\varepsilon \, k \cdot (x_{i} + x_{j+1})
\bigg\}
\, .
\ea
This is finite for $G_{[12] [k k+1]}|_{3 < k < n-1}$, while ${\rm div}_\varepsilon A_{[12][34];0} = {\rm div}_\varepsilon G_{[12][34]}$ (and similarly for mirror 
symmetric graph). It immediately yields the result in Eq.\ \re{SiteConfAnomaly} when subtracted from $A_{[12][23];0}$.

\subsection{Supersymmetric Wilson loop}

Now we are in a position to compute the one-loop anomaly to the Grassmann degree 4 contribution to the supersymmetric Wilson loop 
$\mathcal{W}_{n;1}$. We are not going to restrict ourselves to just the $\chi_1^4$-component, but consider $\mathcal{W}_{n;1}$ in full generality
instead. This requires the inclusion of diagrams with scalars and fermions as well in addition to gauge fields. The only two types of graphs that diverge 
at one loop are the same ones displayed in Fig.\ \ref{ConfAnomaly} with the obvious replacement of gluon propagators by either scalars or 
gauginos. The conformal anomaly at one-loop order for the $\theta^4$ component will be then determined by the sum pairwise contributions, 
identical to the one in Eq.\ \re{PairwiseDecomp}, where now $A = A^{\rm g} + A^{\rm f} + A^{\rm s}$. Let us discuss these in turn.

\subsubsection{Scalars}

Due to its simplicity, let us start with exchanges of scalars. Their link-$[ii+1]$ to link-$[jj+1]$ correlator receives contributions from $E^{[2]}$'s
in \re{connection} and reads
\be
A^{\rm s}_{[ii+1][jj+1];1} = \frac{(ig)^2}{N_c} \int d^D z \, {\rm e}^{i k \cdot z} \VEV{\tr E_{[ii+1]}^{[2]} E_{[jj+1]}^{[2]} \Delta^{\rm s} (z)}
\, ,
\ee
where the scalar insertion $\Delta^{\rm s}$ is given by the last term in Eq.\ \re{DeltaInsertion}. It is instructive to split the calculation into blocks
and first compute the correlation function of $\Delta^{\rm s}$ with scalar fields. Extracting only the terms linear in $k$, one finds after 
taking the integral
\be
\label{ScalarInsertion}
\Delta_{AB,CD} (x_1, x_2) 
=
\int d^D z \, {\rm e}^{i k \cdot z} \VEV{\bar\phi_{AB} (x_1) \Delta_{\rm s} (z) \bar\phi_{CD} (x_2)}
= - \varepsilon_{ABCD} \frac{\Gamma (1 - \varepsilon)}{8 \pi^{2 - \varepsilon}} \frac{k \cdot (x_1 + x_2)}{[- x_{12}^2]^{1 - \varepsilon}}
\, ,
\ee
where the regularized coordinate-space propagators we used reads
$$
\VEV{\bar\phi_{AB} (x_1) \bar\phi_{CD} (x_2)} = \frac{\Gamma (1 - \varepsilon)}{4 \pi^{2 - \varepsilon}} \frac{\varepsilon_{ABCD}}{[- x_{12}^2]^{1 - \varepsilon}}
\, .
$$
Next, inserting Eq.\ \re{ScalarInsertion} into the contour of the supersymmetric Wilson loop, one gets after a lengthy computation for the pole part of the 
divergent Feynman graphs,
\ba
A^{\rm s}_{[12][23];1} 
\!\!\!&=&\!\!\! 
a \frac{\Gamma (1-\varepsilon)}{32 \varepsilon} (- x_{13}^2 \mu^2 \pi)^\varepsilon  \frac{\chi_1 \chi_2}{x_{13}^2 \vev{12}^2}
\\
&\times&\!\!\!
\Big\{
\chi_2 \big(\chi_2 - \vev{12} \eta_1\big) \bra{1} k |2] - \chi_1 \big(\chi_1 - \vev{12} \eta_2\big)  \bra{2} k |1]
- 
\chi_1 \chi_2 \big(\bra{1} k |1] - \bra{2} k |2]\big)
\Big\}
\, , \nonumber
\ea
and\ba
A^{\rm s}_{[12][34];1} 
\!\!\!&=&\!\!\! a \frac{\Gamma (1-\varepsilon)}{32 \varepsilon}
(- x_{13}^2 \mu^2 \pi)^\varepsilon \frac{\chi_1 \chi_2 \chi_3}{\vev{12} \vev{23}}
\left\{
\chi_1 \frac{\bra{2} k |1]}{x_{13}^2}
-
\chi_3 \frac{\bra{2} k |3]}{x_{24}^2}
\right\}
\, ,
\ea
for the nearest- and next-to-nearest-neighbors, respectively.

\subsubsection{Gauginos}

Next, we turn to the contribution of fermions to the anomaly,
\be
A^{\rm f}_{[ii+1][jj+1];1}
= \frac{(ig)^2}{N_c} \tr \, \int d^D z \, {\rm e}^{i k \cdot z}
\VEV{E_{[ii+1]}^{[3]} E_{[jj+1]}^{[1]} \Delta^{\rm f} (z)}
\, ,
\ee
where the insertion $\Delta^{\rm f}$ is now determined by the second term in Eq.\ \re{DeltaInsertion}. Since this insertion is just the fermion 
equations of motion, the calculation can be easily done with propagators in the coordinate representation, i.e., 
$$
\VEV{\psi_\alpha (x_1) \bar\psi_{\dot\alpha} (x_2)} 
= 
i \frac{\Gamma (2 - \varepsilon)}{2 \pi^{2 - \varepsilon}} \frac{(x_{12} )_{\alpha\dot\alpha}}{[- x_{12}^2]^{2 - \varepsilon}}
\, ,
$$
since the insertion $\Delta^{\rm f}$ is effectively moved onto the contour due to the contact nature of the emerging interaction,
$$
\partial^{\dot\beta\alpha}_1 \VEV{\psi_\alpha (x_1) \bar\psi_{\dot\alpha} (x_2)} 
= \delta_{\dot\alpha}^{\dot\beta} \delta^{(D)} (x_{12})
\, ,
$$
and thus leaving just one propagator.  The operator insertion, expanded to linear order in the momentum $k$, takes the form
\be
\Delta_{\alpha\dot\alpha}{}^A{}_B (x_1, x_2)
=
\int d^D z \, {\rm e}^{i k \cdot z}
\VEV{\psi_\alpha^A (x_1) \Delta^{\rm f} (z) \bar\psi_{\dot\alpha B} (x_2)}
=
- \frac{i \Gamma (2 - \varepsilon)}{4 \pi^{2 - \varepsilon}} \, k \cdot (x_1 + x_2)  \frac{x_{12 \, \alpha\dot\alpha}}{[- x_{12}^2]^{2 - \varepsilon}}  \delta^A{}_B
\, .
\ee
Then a calculation of the diagram displayed in Fig.\ \ref{ConfAnomaly} (a) with the gauge boson line being replaced by the fermion one yields
\be
A^{\rm f}_{[12][23];1} = a \frac{\Gamma (1 - \varepsilon)}{48 \varepsilon} (- x_{13}^2 \mu^2 \pi)^\varepsilon 
\frac{\chi_1^2 \chi_2}{x_{13}^2 \vev{12}^2}
\left\{
\chi_1 \bra{2} k |1] + \chi_2 \bra{1} k |1] - \bra{2} k |2] \left( 2 \chi_2 - \vev{12} \eta_1 \right)
\right\}
\, .
\ee
Finally, the exchange diagram of the type given in Fig. \ref{ConfAnomaly} (b) produces
\be
A^{\rm f}_{[12][34];1} = a \frac{\Gamma (1 - \varepsilon)}{48 \varepsilon} (- x_{13}^2 \mu^2 \pi)^\varepsilon 
\frac{\chi_1 \chi_2 \chi_3}{\vev{12} \vev{23}}
\left\{
\chi_2 \frac{\bra{3} k |3]}{x_{24}^2}
-
\chi_1 \frac{\bra{2} k |1]}{x_{13}^2}
\right\}
\, .
\ee  
The contribution with interchanged Grassmann degrees on the sites, i.e., $\VEV{E_{[ii+1]}^{[1]} E_{[jj+1]}^{[3]} \Delta^{\rm f} (z)}$ is determined from the 
above making use of the substitution $1 \leftrightarrow 2$ and flipping the overall sign. Notice that both results are independently translationally invariant 
as the shift-breaking terms, present at the intermediate steps, cancel between contributions of the bosonic and fermionic connections to the individual 
Feynman graphs and serve as cross-check on the calculation.

\subsubsection{Gauge fields}

Finally, we come to the gauge fields which are computationally the most involved contributions. The analysis is best performed in terms of Fourier 
transformed propagators. As in the previous two sections, it is instructive to find the vacuum expectation value of the gluon insertion $\Delta^{\rm g}$ 
with the chiral field strength stemming from the $[i-1i]$-link and gauge fields, from $[i i+1]$ and $[i+1, i+2]$ in Fig.\ \ref{ConfAnomaly} (a) and (b), 
respectively. It reads
\be
\Delta_{\alpha\beta}^{\gamma \dot\gamma} = \int d^D z \, {\rm e}^{i k \cdot z} \VEV{F_{\alpha\beta} (x_1) \Delta^{\rm g} (z) A^{\gamma\dot\gamma}(x_2)}
\, .
\ee
Since $\Delta^{\rm g}$ is a sum of both chiral and antichiral components of the field strength, one needs the following propagators
\baa
\VEV{A_{\alpha\dot\alpha} (x_1) F_{\beta\gamma} (x_2)} 
&=& 
\ft{i}{2} \int \frac{d^D k}{(2 \pi)^D}  \frac{{\rm e}^{- i k \cdot x_{12}}}{k^2} \varepsilon_{\alpha (\beta} k_{\gamma) \dot\alpha}
\, , \\
\VEV{F_{\alpha\beta} (x_1) \bar{F}_{\dot\alpha\dot\beta} (x_2)} 
&=& 
\ft{i}{4} \int \frac{d^D k}{(2 \pi)^D}  \frac{{\rm e}^{- i k \cdot x_{12}}}{k^2} k_{(\alpha \dot\alpha} k_{\beta) \dot\beta}
\, , \\
\VEV{F_{\alpha\beta} (x_1) F_{\gamma\delta} (x_2)} 
&=& 
\ft{i}{4} \int \frac{d^D k}{(2 \pi)^D}  {\rm e}^{- i k \cdot x_{12}} \varepsilon_{(\alpha\gamma} \varepsilon_{\beta)\delta}
\, ,
\eaa
where the braces stand for symmetrization of indices $T_{(\alpha \dots \beta)} \equiv T_{\alpha \dots \beta} + T_{\beta \dots \alpha}$. Then,
the result of a straightforward calculation, keeping only terms linear in $k$, takes the following compact form
\be
\Delta_{\alpha\beta}^{\gamma \dot\gamma} 
=
\frac{i}{4 \pi^{2 - \varepsilon}}
\bigg\{
\left[ 
\ft14 x_{12 (\alpha}^{\dot\gamma} x_{12 \, \beta)\dot\delta} \, k^{\dot\delta\gamma} 
-
k \cdot x_1\,  x_{12 (\alpha}{}^{\dot\gamma} \delta_{\beta)}{}^\gamma 
\right]
\frac{\Gamma(2 - \varepsilon)}{[- x_{12}^2]^{2 - \varepsilon}}
-
\ft14 k_{(\alpha}{}^{\dot\gamma} \delta_{\beta)}{}^\gamma \frac{\Gamma(1 - \varepsilon)}{[- x_{12}^2]^{1 - \varepsilon}}
\bigg\} 
\, .
\ee 
This can be rewritten in the form that one would get from the use of the covariant Lorentz form of the action, i.e., $F_{\mu\nu}^2$. Manipulating the above 
result, one finds
\baa
\Delta_{\alpha\beta}^{\gamma \dot\gamma} 
=
\frac{i}{4 \pi^{2 - \varepsilon}}
\bigg\{
\left[ 
\ft14 x_{12}^{\gamma\dot\gamma} \, x_{12 (\alpha \, \dot\delta} \, k_{\beta)}{}^{\dot\delta} 
-
\ft12 k \cdot (x_1+ x_2) \,  x_{12 (\alpha}{}^{\dot\gamma} \delta_{\beta)}{}^\gamma 
\right]
\frac{\Gamma(2 - \varepsilon)}{[- x_{12}^2]^{2 - \varepsilon}}
-
\ft14 \varepsilon \, k_{(\alpha}{}^{\dot\gamma} \delta_{\beta)}{}^\gamma \frac{\Gamma(1 - \varepsilon)}{[- x_{12}^2]^{1 - \varepsilon}}
\bigg\}
\, .
\eaa
We observe the presence of $O(\varepsilon)$ effects in this form of writing the expression. These are crucial for getting self-consistent results
since the latter contribution induces double poles in $\varepsilon$ in intermediate steps.

The brute force calculation of the diagrams in Fig.\ \ref{ConfAnomaly} is extremely tedious and lengthy. However, the final
result is rather compact. The contribution of the graph \ref{ConfAnomaly} (a) to the anomaly can be written as 
\ba
\label{GluonAnA}
A^{\rm g}_{[12][23];1} 
\!\!\!&=&\!\!\!  
a \frac{\Gamma (1-\varepsilon)}{192 \varepsilon} (- x_{13}^2 \mu^2 \pi)^{\varepsilon} 
\frac{\chi_{1}}{x_{13}^2 \vev{12}^2}
\Big\{
4 \vev{12} \, k \cdot x_2 \, \chi_{1} \chi_2 \eta_{1}
+
( \vev{12}^3 \eta_{1}^3 - \chi_2^3) \bra{1} k | 2 ] 
\\
&&-
\vev{12} \chi_2 \eta_{1}
\Big(
5 \chi_{1} \bra{2} k | 2 ]  - 3 \chi_2 \bra{1} k | 2 ] 
\Big)
+
3 \vev{12}^2 \eta_{1}^2 \Big( \chi_{1} \bra{2} k |2] - \chi_2 \bra{1} k |2] \Big)
\Big\}
\, , \nonumber
\ea
while from (b), it is
\ba
\label{GluonAnB}
&&\!\!\!\!\!\!\!\!
A^{\rm g}_{[12][34];1} 
= 
a \frac{\Gamma (1-\varepsilon)}{192 \varepsilon}
(- x_{13}^2 \mu^2 \pi)^{\varepsilon} \frac{\chi_1 \chi_2}{x_{24}^2 x_{13}^2 \vev{12}}
\bigg\{
-
4 k \cdot x_2 \,x_{24}^2 \,  \chi_{1} \eta_{1}
+ 
\chi_1 \chi_2 [23] \bra{3} k |1]
\\
&&
+
\left(  \vev{23} \chi_{1} - \vev{13} \chi_2 \right)
\bigg[ 
\chi_2 \!
\left[ \frac{[12]}{\vev{12}} \bra{1} k | 3]  
-
\frac{[12]}{\vev{23}} \bra{3} k |3] 
\right] \!
+
\chi_1 \!
\left[
2 \frac{[23]}{\vev{12}} \bra{2} k |1] - \frac{[12]}{\vev{12}} \bra{2} k |3]
\right]
\bigg]
\bigg\}
\, . \nonumber
\ea
Contrary to their scalar/fermion counterparts, both graphs independently receive contribution which are not translationally invariant, i.e., $\sim (k \cdot x_2) $. 
However, these do cancel in the sum of the two, echoing the cancellation observed in the case of the anomaly for the bosonic Wilson loop in Section
\ref{BosonicAnomaly}.

\section{Comments}

Let us comment on the derived expressions. Having found the anomaly in the conformal variation of the super Wilson loop, we can test it against the 
explicit calculations performed in Section \ref{OneLoopSuperW}. Making use of decomposition of $\eta$'s in terms of Grassmann components of 
supertwistors \re{ChiToEta} one can extract the $\chi_1^4$ contribution from Eqs.\ \re{GluonAnA} and \re{GluonAnB}. Adding to these analogous 
terms stemming from the mirror symmetric diagrams, one immediately recognizes the anomalous variation of $\vev{\mathcal{W}}_{n;1}$ displayed 
in Eq.\ \re{Chi4ConfAnom}. Actually, one can go in the other direction and predict the anomalous contribution $a_{n;1}$ to the super Wilson loop by 
solving the conformal Ward identity and fixing a possible additive conformally invariant piece by taking the collinear limit to four points. The latter is purely 
anomalous and was computed at one-loop order for all components of Grassmann degree 4 in Ref.\ \cite{BelKorSok11}. Along this way we can perform a 
subtractive transformation on the super Wilson loop and thus restore the conformal symmetry of the object. It is the latter that is dual to non-MHV amplitudes.

Our final comment addresses the question of what components of the super Wilson loop are anomalous and thus require subtractions. An inspection of 
the right-hand side of the conformal Ward identities immediately suggests that since the only divergent Feynman graphs are those coming from exchanges 
that involve nearest and next-to-nearest links, the degree 4 Grassmann structure can be anomalous provided it contains at most three adjacent indices,
e.g., $\chi_{i-1}^2 \chi_i \chi_{i+1}$, $\chi_{i-1}^2 \chi_{i}^2$ etc. Therefore, any structure where at least one of the indices is not adjacent to the rest will
be conformal and given by the corresponding component of the sum of $R$-invariants.

To conclude, the use of one of the most suitable regularization schemes that preserves the helicity-spinor formalism, one the one hand, and allows to tame
divergences by means of deviation from the four-dimensional space-time, on the other, inevitably breaks conformal symmetry of the supersymmetric 
Wilson loop. However the latter, contrary to correlation functions of gauge invariant operators, is a scheme dependent object. It is accompanied by a
coefficient function it light-cone OPE. So only the product of the two is expected to be anomaly-free. What we suggested here is to subtract the conformally 
non-invariant effects that come from the anomalies in the renormalization of the superloop with the variation of the regularized SYM action. The procedure
is echoing analogous treatment of mixing effects in the renormalization of conformal operators in gauge theories. This implies that the finite subtraction
$a_{n;1}$ should compensate the anomalous effects in the OPE coefficient function computed the same order to yield a purely conformal contribution to
the correlation function of stress-tensor supermultiplet as was found in Ref.\ \cite{EdeHesKorSok11}.

Several directions are open for further investigation. It is important to understand the all-loop structure of the found anomalies. Next, since the consideration
of this work was limited to the NMHV level, it necessary to unravel the conformal anomalies for contribution of higher Grassmann degrees. Last but not least, 
it would be very instructive to define a set of rules on the Lagrangian level that would produce a conformal invariant result after the regularization is removed, 
thus automatically subtracting the notorious conformal anomalies.

\vspace{5mm}

We would like to thank Gregory Korchemsky for helful correspondence and instructive comments on the manuscript and David Skinner for useful discussions.
This work was supported by the U.S. National Science Foundation under grants PHY-0757394 and PHY-1068286.


\end{document}